\documentclass[a4paper,12pt]{article}
\pdfoutput=1
\usepackage{graphicx, rotating,amssymb,amsmath,soul}

\usepackage{jcappub}
\include{aastex_journal_defs}

\ifx\pdfoutput\undefined
\usepackage[dvips,bookmarks]{hyperref}	
\else
\usepackage{hyperref}	
\fi
\hypersetup{colorlinks,bookmarksopen,bookmarksnumbered,citecolor=rossoc,
linkcolor=blue,pdfstartview=FitH,urlcolor=rossos}

\def\hhref#1{\href{http://arxiv.org/abs/#1}{#1}} 

\usepackage{multicol}
\usepackage{color}
\definecolor{rosso}{cmyk}{0,1,1,0.4}
\definecolor{rossos}{cmyk}{0,1,1,0.55}
\definecolor{rossoc}{cmyk}{0,1,1,0.2}
\definecolor{blu}{cmyk}{1,1,0,0.3}
\definecolor{blus}{cmyk}{1,1,0,0.6}
\definecolor{bluc}{cmyk}{1,1,0,0.1}
\definecolor{verde}{cmyk}{0.92,0,0.59,0.25}
\definecolor{verdec}{cmyk}{0.92,0,0.59,0.15}
\definecolor{verdes}{cmyk}{0.92,0,0.59,0.4}

\font\tenrsfs=rsfs10 at 12pt
\font\sevenrsfs=rsfs7
\font\fiversfs=rsfs5
\newfam\rsfsfam
\textfont\rsfsfam=\tenrsfs
\scriptfont\rsfsfam=\sevenrsfs
\scriptscriptfont\rsfsfam=\fiversfs
\def\mathscr#1{{\fam\rsfsfam\relax#1}}

\newcommand{\LM}{L^{\rm max}}
\newcommand{\Lm}{L^{\rm min}}
\newcommand{\LMM}{L_{>100\,{\rm MeV}}^{\rm max}}

\newcommand{\LMG}{L_{>1\,{\rm GeV}}^{\rm max}}
\newcommand{\LmG}{L_{>1\,{\rm GeV}}^{\rm min}}

\def\circa#1{\,\raise.3ex\hbox{$#1$\kern-.75em\lower1ex\hbox{$\sim$}}\,}

\newcommand{\beq}{\begin{equation}}
\newcommand{\eeq}{\end{equation}}

\newcommand{\lsi}{\,\raisebox{-0.13cm}{$\stackrel{\textstyle<}{\textstyle\sim}$}\,}
\newcommand{\gsi}{\,\raisebox{-0.13cm}{$\stackrel{\textstyle> }{\textstyle\sim}$}\,}

\def\circa#1{\,\raise.3ex\hbox{$#1$\kern-.75em\lower1ex\hbox{$\sim$}}\,}
\makeatletter

\def\art{\@ifnextchar[{\eart}{\oart}}
\def\eart[#1]#2#3#4#5#6{{\rm #2}, {#3 #4} {\rm (#6) #5} [{\hhref{#1}}]}

\def\hepart[#1]#2{{\rm #2, \hhref{#1}}}
\newcommand{\oart}[5]{{\rm #1}, {#2 #3} {\rm (#5) #4}}

\newcounter{alphaequation}[equation]

\def\thealphaequation{\theequation\hbox to
0.6em{\hfil\alph{alphaequation}\hfil}}

\def\eqnsystem#1{
\def\@eqnnum{{\rm (\thealphaequation)}}
\def\@@eqncr{\let\@tempa\relax \ifcase\@eqcnt \def\@tempa{& & &} \or
  \def\@tempa{& &}\or \def\@tempa{&}\fi\@tempa
  \if@eqnsw\@eqnnum\refstepcounter{alphaequation}\fi
\global\@eqnswtrue\global\@eqcnt=0\cr}
\refstepcounter{equation} \let\@currentlabel\theequation \def\@tempb{#1}
\ifx\@tempb\empty\else\label{#1}\fi
\refstepcounter{alphaequation}
\let\@currentlabel\thealphaequation
\global\@eqnswtrue\global\@eqcnt=0 \tabskip\@centering\let\\=\@eqncr
$$\halign to \displaywidth\bgroup \@eqnsel\hskip\@centering
$\displaystyle\tabskip\z@{##}$&\global\@eqcnt\@ne
\hskip2\arraycolsep\hfil${##}$\hfil& \global\@eqcnt\tw@\hskip2\arraycolsep
$\displaystyle\tabskip\z@{##}$\hfil
\tabskip\@centering&\llap{##}\tabskip\z@\cr}
\def\endeqnsystem{\@@eqncr\egroup$$\global\@ignoretrue} \makeatother

\def\fermi{\textsl{Fermi}~}
\def\eg{{\em e.g.}\@~}
\def\ie{{\em i.e.}\@~}

\begin{document}

\begin{flushleft}
\scriptsize{
LAPTH-226/14}
\end{flushleft}

\vspace{-0.02cm}

\begin{flushleft}

{\Huge\bf Millisecond pulsars and the Galactic Center gamma-ray excess: the importance of luminosity function and secondary emission}

\bigskip
\bigskip\color{black}\vspace{0.4cm}

{
{\large\bf Jovana Petrovi\'c}$^{\,a,b}$,
{\large\bf Pasquale D. Serpico}$^{\,c}$,
{\large\bf Gabrijela Zaharijas}$^{\,d,e,f}$
}
\\[7mm]

{\it $^a$ {\href{http://astro.matf.bg.ac.rs/beta/index.php?lang=eng}{Department of Astronomy},  Faculty of Mathematics, University of Belgrade, Studentski trg 16 , 11000 Beograd, Serbia}}\\[3mm]
{\it $^b$ {\href{http://www.pmf.uns.ac.rs/en/about_us/departments/physics}{Department of Physics}, Faculty of Sciences, University of Novi Sad, Trg Dositeja Obradovi\'ca 4, 21000 Novi Sad, Serbia}}\\[3mm]
{\it $^c$ Laboratoire de Physique Th{\'e}orique d' Annecy-le-Vieux (\href{http://lapth.cnrs.fr/}{LAPTh}), Univ. de Savoie, CNRS, B.P.110, Annecy-le-Vieux F-74941, France}\\[3mm]
{\it $^d$ Laboratory for Astroparticle Physics (\href{http://www.ung.si/en/research/laboratory-for-astroparticle-physics/}{LAPP}),\\
	University of Nova Gorica, Vipavska 13, SI-5000 Nova Gorica, Slovenia}\\[3mm]
{\it $^e$ Abdus Salam International Centre for Theoretical Physics (\href{http://www.ictp.it}{ICTP}),
	Strada Costiera 11, 34151 Trieste, Italy}\\[3mm]
{\it $^f$ Istituto Nazionale di Fisica Nucleare - Sezione Trieste (\href{http://www.ts.infn.it/}{INFN}),\\
		Padriciano 99, I - 34149 Trieste, Italy}

\bigskip		
{\bf E-mail:} jovana.petrovic@df.uns.ac.rs, serpico@lapth.cnrs.fr, gabrijela.zaharijas@ung.si		

\end{flushleft}

\newpage
\begin{quote}
\color{black}
\large
{\bf Abstract: } Several groups of authors have analyzed \fermi LAT data in a region around the Galactic Center finding an unaccounted gamma-ray excess over diffuse backgrounds in the GeV energy range. {It has been argued that it is difficult or even impossible to explain this diffuse emission by the leading astrophysical candidates - millisecond pulsars (MSPs).} Here we {provide a new estimate of the contribution to the excess by a} population of yet unresolved MSP located in the bulge of the Milky Way. {We simulate this population with the {\sc GALPLOT} package by adopting a parametric approach, with the range of free parameters gauged on the MSP characteristics reported by the {second pulsar} catalogue  (2PC)}. We find that the conclusions strongly depend on the details of the MSP luminosity function {(in particular, its high luminosity end)} {and other explicit or tacit assumptions on the MSP statistical properties, which we discuss. Notably, for the first time we} {study} {the importance of} the { possible secondary emission of the MSPs in the Galactic Center}, {i.e. the emission via inverse Compton losses of electrons injected in the interstellar medium.  Differently from} {a majority of} {other authors, we find that} within current uncertainties a large if not dominant contribution { of MSPs} to the excess cannot be excluded. We also show that {the sensitivities of future instruments or possibly already of the latest  LAT data analysis (Pass 8) provide  good perspectives to test this scenario by resolving a significant number of MSPs.}
\end{quote}

\section{Introduction} \label{sec:intro}

{The Galactic center (GC) is a complex yet interesting region which has been and still is in the focus of astro-particle research.  For instance, it is expected to harbor large quantities of dark matter (DM) particles that---at least in  Weakly Interacting Massive Particle (WIMP) models---can produce significant annihilation emission. Furthermore, the GC also hosts the closest supermassive black hole and a variety of non-thermal astrophysical sources, making this environment difficult to model.}

Not unexpectedly, many research groups  have analyzed \fermi {Large Area Telescope (LAT)} data in the GC region motivated by the prospect of detecting DM signatures. Excitingly, several groups \cite{Abazajian:2014fta,Daylan:2014rsa,Hooper:2011ti,Hooper:2010mq,Hooper:2013rwa,Boyarsky:2010dr,Abazajian:2012pn,Gordon:2013vta,Calore:2014xka} claimed a detection of a GeV gamma ray excess, above a model of the expected astrophysical emission. {Analyses performed within \fermi LAT collaboration had longstanding hints of this excess (see e.g.~\cite{Vitale:2009hr}),  consistent with preliminary results of the 5 year data analysis presented recently\footnote{http://fermi.gsfc.nasa.gov/science/mtgs/symposia/2014/program/08\_Murgia.pdf}. {Most works suggest similar key} properties of this residual emission (referred to as Galactic Center Excess (GCE) in what follows): i) its quasi-spherical shape, with spatial extension declining as $\sim r^{-2.4}$ up to about $10^\circ$ away from the GC; ii) its power law spectral shape with an exponential cut-off  $E^{-\Gamma} \text{exp} [-E/E_{\text{cut}}]$, whose parameters are in the range $\Gamma=1.6\pm 0.2$ and $E_{\text{cut}}\sim 4.0\pm 1.5$ GeV, \cite{Gordon:2013vta}, iii) its total flux of roughly $10^{-7} $ ph cm$^{-2}$ s$^{-1}$ in the $7^\circ \times 7^\circ$ region centered on the Galactic center \cite{Gordon:2013vta}. The spectral shape as well as the morphology of the GCE could be fit by the annihilation of a WIMP of $\sim30\,$GeV mass into $b\bar{b}$ particles, as suggested by several groups~\cite{Hooper:2011ti,Hooper:2010mq,Hooper:2013rwa,Huang:2013apa,Abazajian:2012pn,Calore:2014xka}.

As alluring as the idea of discovering a DM signal might be, {a few caveats should be kept in mind:} the publicly available \fermi  diffuse model \footnote{http://fermi.gsfc.nasa.gov/ssc/data/access/lat/BackgroundModels.html}, used in a majority of the studies suggesting a DM interpretation of the GCE, is in fact fitted to data for the main purpose of studying point sources and is thus not optimal to {isolate an extended signal, even less so in presence of poorly know foreground sources}. Only recently there have been efforts to start addressing the systematic uncertainties in the derivation of the residual emission, showing the high model dependence of the signal especially below 1 GeV \cite{Gordon:2013vta,Abazajian:2014fta}. In particular,  the recent investigation \cite{Calore:2014xka} suggests quantitative differences with the above results once a more extended set of background models is allowed for, {noticing that the spectra of the GCE might extend to higher energies and that a broken power-law can fit the excess.} { The possibility that a non-negligible fraction of the excess~(\cite{Calore:2014xka}, see also the talk in footnote 1) or its totality~\cite{Gaggero:2014xla, Mirabal:2014ila}  can be explained by a more sophisticated modeling of the conventional diffuse emission may still be open.}

{It is also mandatory to carefully} investigate alternative astrophysical explanations for the gamma ray GCE. Millisecond pulsars (MSP) have been considered as promising candidates in~\cite{Abazajian:2012pn,Yuan:2014rca,Mirabal:2013rba}, alongside recently proposed secondary emission from {Cosmic Ray (CR)} protons~\cite{Carlson:2014cwa} or electrons~\cite{Petrovic:2014uda} injected in a vicinity of the GC, about a Myr ago. The appeal of MSPs is due to a number of reasons: First, typical MSP spectra have both the power law index and the energy cut-off intriguingly similar to those of the GCE. Second, a dense GC environment is known to have hosted several star burst episodes in the past \cite{Ponti:2012pn} and could therefore naturally account for a large number of binary systems needed for MSP formation. {In addition, a number of arguments supporting the possibility that MSPs in the bulge are consistent with a relatively extended distribution---closer to observed GCE  declining as $\sim r^{-2.4}$---have been proposed in the literature, see~\cite{Abazajian:2012pn} and refs. therein.} 

The MSP explanation for the origin of the GCE has been studied in a series of papers~\cite{Hooper:2013nhl,Yuan:2014rca,Cholis:2014lta,Cholis:2014noa}. {Based on the observed properties of the MSP population, the authors of these articles have investigated how well MSP spectral properties match the energy distribution of the GCE and whether the MSP luminosity function permits a sufficient number of unresolved MSPs to explain the diffuse GCE, while not over-predicting the number of the observed number of MSPs from given regions.  In several cases, the viability of MSP as an explanation has been excluded or severely challenged, see notably~\cite{Hooper:2013nhl,Cholis:2014lta}.}

{Given the importance of this topic, we decided to revisit the issue in this article, by postulating the existence of a population of  MSPs hosted in the inner region of our Galaxy, \ie extending to $\lsi$ 2 kpc around the GC.}
{In our study of the MSP luminosity function} {we rely on current knowledge on the Galactic MSP population to establish the overall framework of our analysis and benchmark parameters: in particular, we refer to the 2nd Pulsar Catalogue (2PC) \cite{TheFermi-LAT:2013ssa}, as well as to the dedicated study  presented in \cite{Cholis:2014noa}.
The main novelty of our treatment is that we try to limit the number of theoretical assumptions to a minimum and stick to a phenomenological approach by allowing
the key parameters to vary in a reasonably large range of values. We also consider possible systematic effects which may be of relatively minor importance in the earlier MSP population studies focused e.g. on the local Galactic environment, but may play a more significant role in the GC: this is for instance the case of the endpoint of the luminosity function, which does not alter dramatically the conclusions drawn on the properties of the bulk of the local MSP population. 

 {Concerning the agreement between GCE and the average MSP spectra, most of the works focused on the discrepancy in the 100 MeV to $\sim$ 1 GeV range, at which the MSP spectra appears somewhat softer than that of the GCE. However, as it was shown that this discrepancy is within the systematic uncertainty  \cite{Gordon:2013vta,Abazajian:2014fta,Calore:2014xka} which is large in that energy range, we do not address this issue here. A more recent work \cite{Calore:2014xka}, however, found that the GCE spectra extends to energies above the MSP cut-off. In order to explore this high energy spectral discrepancy, we also study (to the best of our knowledge, for the first time in this context) the possible role played by the gamma-ray emission associated to energetic $e^\pm$ released by the MSPs in interstellar space.} We dub these photons ``secondary'' as opposed to the ``prompt'' or ``primary'' photons associated to emission mechanisms in the magnetosphere. This secondary inverse Compton emission has some peculiarities related to the high density and roughly spherical distribution of the interstellar light in the inner Galaxy. 
Our study is thus complementary to other articles  (like~\cite{Cholis:2014noa}) which try to deduce MSP population properties; we rather focus 
on the characteristics that a putative MSP population should have if it is to explain the GCE, and examine if this is consistent with the lack of identification of resolved MSPs in given regions around the GC. We also pay attention to potential signatures which would allow to test this scenario, notably the expectations for forthcoming observations characterized by improved  point spread function and therefore point source sensitivity.}

This article is structured as follows: {In Sec.~\ref{formalism}, after some generalities on MSPs, we devote three subsections to: i) introduce the key parameters entering our description of the putative additional MSP population, with some notions recalling the known properties of ``prompt'' high-energy emission of MSPs, including crucial uncertainties (subsec.~\ref{knownMSP}). ii)  outline some more technical aspects concerning the choice of our region of interest and our implementation in GALPLOT (subsec.~\ref{bulgeMSP}); iii) present and discuss our results (subsec.~\ref{sec:results}).} Section~\ref{ICmsps} discusses inverse Compton signals coming from electrons injected by the same population of the MSPs. Section~\ref{sec:conclusions} is devoted to our conclusions, as well
as perspectives for the future.


\section{Prompt gamma emission from MSPs }\label{formalism}

{Millisecond pulsars are empirically defined as  pulsars with a rotational period in the range of about 1-10 milliseconds. They} are usually {interpreted and} referred to as recycled pulsars due to their evolution from ordinary to millisecond ones in binary systems,  in which MSPs are spun up by accreting the material from their companion. The GC is expected to harbor a vast number of MSPs because of the large numbers of binaries populating this region (for more info about their emission mechanisms and possible sub-populations see \eg \cite{Johnson:2012rk}). 
\fermi LAT has been exceptionally good in studying this new class of gamma ray pulsars: 40 were observed in {three} year data~\cite{TheFermi-LAT:2013ssa}, while the recent work \cite{Cholis:2014noa} reports 61 MSPs within the 5.6 year dataset, and the discovery of 4 new MSPs with Pass 8 has been reported at the Fermi Symposium\footnote{http://fermi.gsfc.nasa.gov/science/mtgs/symposia/2014/abstracts/135}.
Due to their low luminosity, MSPs have only been observed either in our `Galactic neighborhood' (a majority of objects lies within $\sim 3\,$kpc from the Sun with the farthest one$\,\lesssim 6$ kpc away) or in globular clusters, which are known to harbor high densities of MSPs. In globular clusters only two pulsars have been individually resolved (PSR J1823-3021A in NGC 6624 and PSR J1824-2452A in M28), due to a comparatively large angular resolution of the LAT. The  spectra of globular clusters are thus interpreted as the total emission of a large population of unresolved MSPs hosted in these sites.
Due to the significant numbers of MSPs observed either individually or in a global way in globular clusters, it has recently become possible to deduce their average population properties, in particular their average spectrum and their luminosity function, see for instance~\cite{Caraveo:2013lra, Gregoire:2013yta,Hooper:2013nhl}. {These results will be used to benchmark or parameterize the conjectured bulge MSP gamma-ray population and its key parameters.}

\subsection{{Bulge gamma-ray MSP population: parameterization and key inputs}}\label{knownMSP}

\begin{itemize}
\item {\bf Spectra}: {We parameterize the energy spectrum as a power law with an exponential cut-off}
\begin{equation}
\frac{d\Phi_{\gamma}}{dE}\propto \left( \frac{E}{E_{0,\gamma}}\right)^{-\Gamma_{\gamma}} \exp\left(-\frac{E}{E_{\rm cut, \gamma}}\right) \label{eq:spectra}\,.
\end{equation}
{This functional form is  directly taken from analyses of observed MSP spectra and in fact is very similar to the fits of the GCE spectrum.
To get an idea of the values of the parameters needed to fit the GCE, the analysis in \cite{Gordon:2013vta}  found $E_0=1.2 $ GeV, $E_{\rm cut}=4(\pm 1.5)$ GeV and $\Gamma=1.6\pm 0.2$. The more recent analysis~\cite{Calore:2014xka} found that the best fit parameters of the above function are $\Gamma=0.945^{+0.36} _{-0.5}$ and $E_{\rm cut}=2.53^{+0.11} _{-0.77}$ GeV. 
For comparison, the fit  to the observed MSP reported in \cite{Cholis:2014noa} ($E_0=1 $ GeV, $\Gamma=1.57^{+0.01} _{-0.02}$, and $E_{\rm cut}=3.78^{+0.15} _{-0.08}$ GeV)  are in  agreement  within 2 $\sigma$ with} {the measured GCE spectrum}. 

For simplicity, we stick here to the best fit parameters from~\cite{Cholis:2014noa}, given the reasonable agreement among predicted MSP spectra and those derived from the residuals, and focus instead on the more pressing issue of the MSP detectability. {Note that keeping the spectrum fixed also allows one to use interchangeably the gamma-ray energy luminosity quantities (say erg$\,{\rm s}^{-1}$ or erg$\,{\rm cm}^{-2}\,{\rm s}^{-1}$) or the photon number luminosity quantities (say, ph$\,{\rm s}^{-1}$ or ph$\,{\rm cm}^{-2}\,{\rm s}^{-1}$), since they are linked univocally. Fixing the spectrum is not expected to alter dramatically the conclusions. If anything, by allowing for some variation of the spectral parameters would enlarge the range of predictions and thus provide further margins for the MSPs to account for the GCE.}

\item {\bf {Shape of the gamma-ray luminosity} function}: {We assume that the distribution of gamma-ray sources as a function of their gamma luminosity, i.e. the gamma-ray luminosity function, can be approximated with a power-law in the range of luminosities $\Lm$ to $\LM$ of importance for the problem at hand. We shall discuss the pertinence of this approximation  below. We devote the next bullet item to the discussion of $\Lm$ and $\LM$, while dealing here with 
 the only parameter describing the shape (apart for the normalization), namely  the power-law index $\alpha_L$} defined through the equation

\begin{equation}
\frac{dN}{dL_\gamma}\propto L_\gamma^{-\alpha_L} \label{eq:alpha}\,.
\end{equation}
{We do not impose a specific choice of $\alpha_L$, but test three different values in a range reasonably large to include results at the same time consistent with local observations
and some theoretical expectations, whose rationale we now briefly discuss.}

{Observations  suggest the approximate correlation $L_{\gamma}\propto {\dot E}^{\delta}\,,$ where $ {\dot E}$ is the spin-down power, which in turn hints to a power-law dependence of the luminosity function
from the spin-down energy. The measured $L_{\gamma}-{\dot E}$ correlation for MSPs indicates $\delta \sim (0.5 \-- 1)$~\cite{Johnson:2012rk,2010ApJS..187..460A,TheFermi-LAT:2013ssa}, with the recent~\cite{Venter:2014zea} arguing that a value $\delta \sim 1$ is favored for MSP data and is also compatible with expectations  for such old objects.}
{In principle, different models of magnetospheric emissions, if supplemented by some additional assumptions (see e.g.~\cite{Strong:2006hf} for such an early toy model) may  predict the expected value of $\alpha_L$, with ``reasonable'' values for $\alpha_L$ in the range of one to more than two. }

One can also use the {\it Fermi} LAT to {\it derive} the luminosity function, provided that a complete sample exists for some range of luminosities. In \cite{Cholis:2014noa},  the authors derived the luminosity function in two independent ways: i) starting from the luminosity function of the observed nearby MSPs and correcting for the `completeness' factor due to the {\it Fermi} LAT point source sensitivity; and ii) focusing on the gamma ray emission from 11 globular clusters, originating in the MSPs and using a correlation with X-ray luminosity of MSPs in those globular clusters. 
In the range of luminosities in which the pulsar numbers are large enough to be statistically significant {\it and} the sample of field MSPs is almost complete (roughly $10^{33}- 10^{34}$ erg s$^{-1}$) the authors of \cite{Cholis:2014noa} find that the function $L\,dN/dL$ is approximately flat, suggesting $\alpha_L \simeq 1$. { Differently, a broader range of values $\alpha_L \in [0.7;\, 1.5]$ is argued by the authors of \cite{Yuan:2014rca} to reproduce the data at low luminosities, while the high-luminosity range is characterized by a softer luminosity function}. Since near the cut-off the statistics of MSPs is still quite scarce, and since the complete MSP sample spans a narrow range of luminosities, we will explore the consequences of assuming the three values $\alpha_L = 1, 1.5, 2$, which seem relatively close to what is needed to account for observations. 

We stress that for the problem of interest here, a  description of the MSPs luminosity function over the whole domain of $L_\gamma$ is actually unnecessary to a large extent{, a point that has not always been clearly stated in past}: the results depend mostly on the luminosity function in the last decade below the high luminosity cut-off, as we shall argue when commenting Table~\ref{tab:ratioflux}. Therefore we believe that a single power-law modeling of the luminosity function and the relative freedom we leave to the spectral index are well justified. At the moment we do not consider the possibility of a break in the luminosity function, in order to minimize the number of free parameters, but it is worth keeping in mind that the very parameterization used might be an oversimplification of the phenomena of interest, and such a break may be in fact needed especially for the steeper choice $\alpha_L \sim 2$. Nonetheless, constraints on the MSP capability of explaining the GCE based on fitting MSP population properties at lower luminosities and extrapolating them to the region of interest {(as sometimes done in past literature)} should be taken with a grain of salt.

\item {\bf Maximal and minimal {gamma-ray} luminosity:} 
The recent work~\cite{Cholis:2014noa} used the positions of $61$ pulsars observed in gamma rays in multi-wavelength campaigns (and summarized on \fermi  LAT public pages \footnote{https://confluence.slac.stanford.edu/display/GLAMCOG/Public+List+of+LAT-Detected+Gamma-Ray+Pulsars}) to derive their spectral fluxes, analyzing 5.6 yrs of the LAT data.  Then, the authors of~\cite{Cholis:2014noa} calculated the intrinsic luminosities of these pulsars based on their distances reported in the ATNF catalog, and used these numbers to infer the intrinsic luminosity range and luminosity function of MSPs. The minimal and maximal {gamma-ray} luminosities for the MSPs observed in the Solar neighborhood thus obtained are $L_{>100\, {\rm MeV}}^{\rm min}=6\times 10^{31}$ erg~s$^{-1}$ and $L_{>100\, {\rm MeV}}^{\rm max}=2\times 10^{35}$ erg~s$^{-1}$, respectively.

It is important to notice that $L_{>100 {\rm MeV}}$=$1\times 10^{35}$ erg~s$^{-1}$ at 8 kpc corresponds to a flux {$\Phi_{>100 {\rm MeV}}=1.3 \times 10^{-11}$ erg~cm$^{-2}$~s$^{-1}$ (or equivalently, $\Phi_{>1 {\rm GeV}}\simeq 2\times 10^{-9}$ ph~cm$^{-2}$~s$^{-1}$)},  barely above the point source sensitivity threshold for sources with pulsar like spectrum of the \fermi  LAT at $\leq 10^\circ$ latitudes \cite{TheFermi-LAT:2013ssa}. One can thus anticipate---and we shall confirm that in our numerical study reported below---that the analysis is {\it crucially dependent} on the shape of the luminosity function close to the luminosity cut-off, as well as to the value of the cut-off itself. 
For that reason we pay a special care to the choice of $L^{\rm max}$ in this work. The most luminous pulsar {in gamma rays} in the catalogue used by~\cite{Cholis:2014noa} is J0218+4232, which thus sets $L^{\rm max}$. In the ATNF catalog this object was estimated to be at a distance of $5.6$ kpc, implying that its intrinsic luminosity is $2\times 10^{35}$ erg~s$^{-1}$. In \cite{Verbiest:2014xea} this distance estimate was challenged and it was argued that when distance estimates are properly corrected,  a gamma-ray luminosity of $5.4 \times 10^{34}$ erg s$^{-1}$ is obtained, in line with values obtained for other MSPs.  
Also, the second most luminous pulsar listed in~\cite{Cholis:2014noa} has a distance reported in the 2PC catalog which is a factor of 1.6 smaller than the one reported in ATNF. Its luminosity could be thus be a factor of 2.5 times lower than the reference 1.1 $\times 10^{35}$ erg s$^{-1}$.  Note that even in the study in \cite{Calore:2014oga} based on radio properties of MSPs there is {\it no} MSP with a luminosity higher than  $\LMM=10^{35}$ erg s$^{-1}$ (see their Fig. 6, where the highest luminosity pulsar has mean luminosity of $\sim5 \times 10^{34}$ erg s$^{-1}$). {Similar conclusions can be drawn from~\cite{Venter:2014zea} (see in particular their Fig. 3), where
a number of models is used to deduce the {\it beaming factor}---i.e. the fraction of the total solid angle that is swept by the pulsar beam in one rotation---and thus $L_\gamma$ from the observed data: in all cases, $\LMM$ is found to lie below $10^{35}$ erg~s$^{-1}$.}
Note also that while a nominal error on pulsar distances is $\sim 30\%$, distance uncertainty estimates based on dispersion measurements suggest that for about $75\%$ of the directions in the sky the accuracy is no better than a factor of 1.5 to 2~\cite{Schnitzeler:2012jq,Hou:2014haa}.

This illustrates the fact that the $ L^{\rm max}$ parameter, which is critical for observational prospects of MSPs in the GC region, bares substantial uncertainty. To gauge its impact, we will chose three representative values of $\LMG$, shown in Table \ref{tab1}. {The exact value of  $ L^{\rm min}$ is instead much less important, for reasons that we will illustrate when discussing our results in Sec.~\ref{sec:results}, and has thus been kept fixed.}

\end{itemize}

\begin{table}[t]
\begin{center}
\begin{tabular}{|c|c| }
\hline

\textbf{Parameter}  &{  \textbf{Value(s)} } \\
\hline \hline
$\alpha_L$&   1, 1.5, 2 \\
$\LMG$ [ph s$^{-1}$] &  $(1,~1.5,~3)\times 10^{37}$  \\
$\LmG$ [ph s$^{-1}$] &  $6\times  10^{34}$ \\
\hline
$\alpha_R$&    2.4 \\
$\beta_R$ [kpc] &   1 \\
\hline
$\Gamma_{\gamma}$&    1.57 \\
$E_{\rm cut,\gamma}$ [GeV]&    3.78 \\
\hline
\end{tabular}
\caption{Values of benchmark parameters: $\alpha_L$ is the index of the luminosity function defined in Eq. \ref{eq:alpha}; $\LMG$ ($\LmG$) is the maximal (minimal) luminosity of the MSP population; $\alpha_R$ and $\beta_R$ describe the spatial distribution of MSPs, as in Eq.~(\ref{eq:rho});   $\Gamma$ and $E_{\rm cut}$ describe the gamma-ray spectra, see Eq.~(\ref{eq:spectra}).}
\label{tab1}
\end{center}
\end{table}

\subsection{Simulation of the MSP bulge population}\label{bulgeMSP}
{Within the range of} parameters outlined above, we simulated a population of MSPs in the bulge of our Galaxy. The main questions we aim to address are: i) for which set of MSP parameters the GCE can be explained by a population of unresolved MSPs, ii) in those cases, how many MSPs should have been observed by \fermi  LAT and iii) how many pulsars {\it will be} observed with an instrument with point source sensitivity a factor a few times better than the \fermi  LAT (or even with the upcoming Pass 8 {\it Fermi} LAT data which will have an improved angular resolution \cite{2014AAS...22325603G}). 
{Monte Carlo} simulations of MSPs have been performed with the {\sc GALPROP}\footnote{ See http://sourceforge.net/projects/galprop/ and \cite{2011CoPhC.182.1156V}.} plotting package {\sc GALPLOT}\footnote{ http://sourceforge.net/projects/galplot/}, which also includes a source population synthesis part. 

In order to choose a region of interest, we compare the measured cumulative flux of the residuals in various ROIs presented in literature with the {\it Fermi} LAT point source sensitivity in each, see Table \ref{tab2}. We approximate the {\it Fermi} LAT sensitivity with a {\it constant} value within each ROI, however we will choose ROIs small enough such that this approximation holds. 
The authors of~\cite{Calore:2014xka} determined the extension of the GCE concluding that it reaches {\it at least} up to $1\,$kpc (with 2$\sigma$ confidence). 
This is consistent with the results reported in~\cite{Daylan:2014rsa}, which show that a residual emission is present in $1^\circ$ ring templates out to $\sim 12^\circ$, beyond which it becomes difficult to discern it due to increasing  statistical error. In light of these results, we focus here on the regions below 10$^\circ$.
From the Table \ref{tab2} it is clear that both the total fluxes in the three ROIs and the \fermi  LAT point source sensitivity are comparable, and we chose the second row ROI, (dubbed ROI I+II in~\cite{Calore:2014xka}) to present our results.  {Note that in the past (see for instance~\cite{Hooper:2013nhl}), constraints have also been derived by focusing on the relatively high latitude region $|b|>10^\circ$, where the \fermi  LAT threshold is lower. The recent study~\cite{Calore:2014xka} suggests however that at such high distances the excess might be consistent with zero at 95\% C.L. within systematics, hence the need for a sizable MSP  population and the corresponding constraints are less robust.}

Technically, the input parameters in the {\sc GALPLOT} code are:
\begin{itemize}
\item {\bf spatial distribution:} {\sc GALPLOT} is pre-equipped with a spatial distribution function for MSPs, but we choose to introduce some changes in order to reproduce the spatial characteristics of the GCE, \ie the steeply falling function of the distance $r$ from the GC $\sim r^{-2.4}$. In terms of cylindrical coordinates ($z$ and $R$) used in {\sc GALPLOT}, the source density is given as:

\begin{equation}
\rho (r=\sqrt{z^2+R^2}) =  r^{- \alpha _R} \exp \left( -\frac{r}{\beta _R} \right) \label{eq:rho}
\end{equation}

 \item {\bf luminosity function:} we set $L^{\rm min}$, $L^{\rm max}$ and the index of the luminosity function $\alpha_L$ to the values defined above and summarized in Table \ref{tab1}. We then {vary 
 $L^{\rm max}$  and $\alpha_L$} to show their impact on the results.   
 For every set of parameters we renormalize the total number of MSP in the bulge so that our unresolved MSP population matches the {best-fit} flux of the GCE in our region of interest, see Table \ref{tab2}. 
 
\item {\bf energy spectrum:}  since this is probably the least controversial point, in this work we did not aim at an optimal  match of the energy spectrum; hence we simply use {the functional form of Eq.~(\ref{eq:spectra})
with parameters chosen according to the discussion in the first bullet in sec.~\ref{knownMSP}}.   
\end{itemize}

\begin{table}[t]
\begin{center}
\begin{tabular}{|c|c|c|c| }
\hline

\textbf{ROI}  & \textbf{F}$_{\rm GCE}$  & \textbf{F}$_{\rm PS~sensit.}$ & \textbf{Ref.}  \\
\hline \hline
$  |b|\leq 3.5^\circ, |l|\leq 3.5^\circ $&   $ 1.~ ({0.5/1.7}) ~ 10^{-7}$ &  $1.5  ~ 10^{-9}$& \cite{Gordon:2013vta} \\
${\bf  \sqrt{b^2+l^2}\leq 5^\circ, ~|b|\geq |l|,~ |b|\geq 2^\circ} $&   ${\bf 0.6 ~(0.4/0.75) ~ 10^{-7}}$ & ${\bf 1.3  ~ 10^{-9}}$& \cite{Calore:2014xka} \\
$ 5^\circ \leq \sqrt{b^2+l^2}\leq 10^\circ, ~|b|\geq |l|,~|b|\geq 2^\circ $&   $0.45 ~(0.25/0.65) ~ 10^{-7}$ & $1. ~ 10^{-9}$& \cite{Calore:2014xka} \\
$ 10^\circ \leq \sqrt{b^2+l^2}\leq 15^\circ, ~|b|\geq |l|,~|b|\geq 2^\circ $&   $0.3 ~({0.1/0.5}) ~ 10^{-7}$ & $0.8 ~ 10^{-9}$& \cite{Calore:2014xka} \\

\hline
\end{tabular}
\caption{Fluxes of the GCE from different regions of interest, compared with \fermi LAT point-source sensitivity. Fluxes are calculated above 1 GeV assuming a pulsar like spectrum {(see first bullet in sec.~\ref{knownMSP})} and expressed in {[ph cm$^{-2}$ s$^{-1}$]}. For F$_{\rm GCE}$ we quote the middle value, with min/max flux values in parentheses.}
\label{tab2}
\end{center}
\end{table}



\subsection{Results}
\label{sec:results}

\begin{figure}[t]
\begin{center}
\includegraphics[width=.99\textwidth]{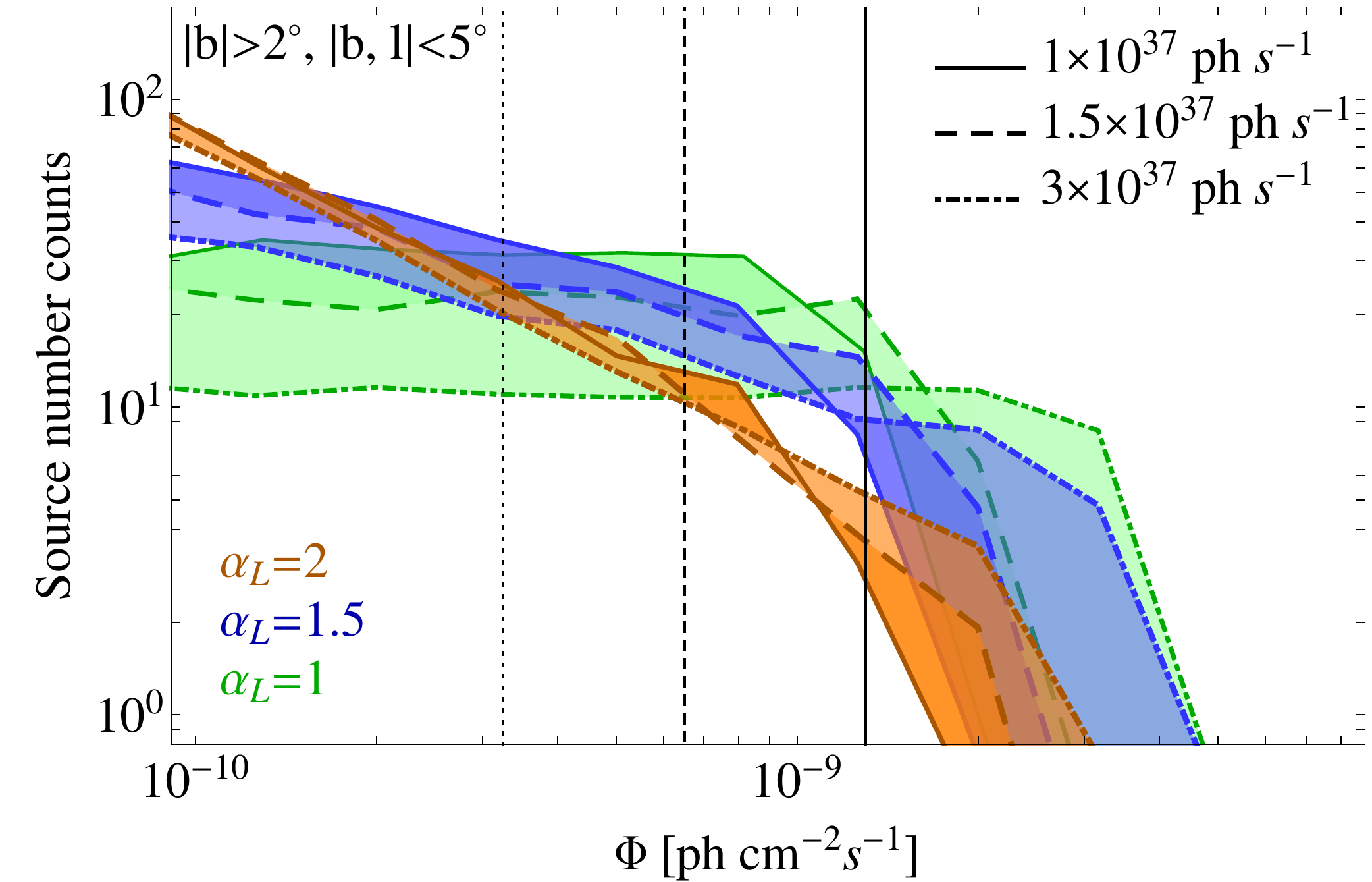}
\caption{Numbers of simulated MSPs in our regions of interest (${ \sqrt{b^2+l^2}\leq 5^\circ, |b|\geq 2^\circ}$ ) in 0.2 logarithmic size gamma-ray flux bins, for parameter values spanning our parameter space: we explore the values of the maximal luminosity of MSP, $L_{\rm max}=1.~(1.5, ~3.)\times 10^{37}$ [ph s$^{-1}$] shown with {({\it Solid})} ({\it Long-Dashed}, {\it Dot-Dashed}) lines. Each color band shows results for three assumptions on the luminosity index $\alpha_L$ varying between $2$ ({\it Orange}), $1.5$ ({\it Blue}) and $1$ ({\it Green}), shown respectively from top to bottom, on the left of the figure. Note that the lines corresponding to the same value of $\alpha_L$ (same color band) but different $L_{\rm max}$ (linestyles) are intersecting, as indicated by the different intensity of color shading. Vertical lines show the current {\it Fermi} LAT point source sensitivity in our ROI taken from \cite{TheFermi-LAT:2013ssa} ({\it Solid} line) in addition to two times improved ({\it Dashed}) and four times improved ({\it Dotted} line) sensitivity. \label{fig:benchmark}}
\end{center}
\end{figure}

{ In Fig.~\ref{fig:benchmark} we show with different line styles the source counts for three luminosity functions corresponding to the choices of $L^{\rm max}$ reported in Table~\ref{tab1}, {normalized in such a way that their overall flux above $\LmG$ matches the central value for the GCE. Each color band, instead corresponds to a different value of $\alpha _L$.}} The key numerical results corresponding to this plot are also reported in Tab.~\ref{tab:numbers}. { In this figure,} we zoom on the high luminosity end of the function which is the most relevant for our discussion. In fact, when setting
$\LmG$ to the value reported in Tab. \ref{tab1} (motivated by observation of nearby faint objects),  sources with fluxes $\gsi 10^{-10}$ ph cm$^{-2}$ s$^{-1}$ (corresponding roughly to $L_{>0.1{\rm GeV}}=5\times 10^{33}$ erg s$^{-1}$),  contribute to the majority of the signal, as reported in Table \ref{tab:ratioflux}. Being the lowest allowed normalization up to a factor two smaller than the central value (see Table~\ref{tab2}), this means that the behavior below this threshold is barely relevant for the problem at hand.

The rightmost (solid) vertical lines in each panel report the {\it Fermi} LAT threshold sensitivity in the shown ROI, as given in the {second} \fermi LAT pulsar catalog (see section 8.2 in~\cite{TheFermi-LAT:2013ssa}). 
The middle (dashed) and leftmost (dotted) vertical lines present twice and four times improved sensitivity, respectively. Indicatively, the sensitivity marked by a dashed line might be already achieved with the forthcoming {\it Fermi} LAT (Pass 8) event selection, while further significant improvements in this region of the sky as marked by the dotted line should probably wait for the next generation gamma-ray space experiments~\cite{Wu:2014tya,Galper:2013sfa}.

\begin{table}[t]
\begin{center}
\begin{tabular}{|c||c|c|c|c|c| }
\hline

 $L^{\rm max}$     [ph s$^{-1}$]                   & $\alpha_L=1$                 & $\alpha_L=1.5$                  & $\alpha_L=2$  \\
\hline \hline
$1\times 10^{37}$&                                          2.7 (45/93)           &          1.6 (30/75)               &    0.6 (14/40)   \\
$1.5\times 10^{37}$&                                      15 (48/83)             &         10 (35/70)                &     3.4 (13/41)  \\
 $3\times 10^{37}$ &                                          25 (42/58)             &        17 (35/62)              & 6.6 (18/43)      \\

\hline
\end{tabular}
\caption{The predicted {\it average} numbers of resolved MSPs for {\it Fermi} LAT point source sensitivity for the full parameter space of our parameters. Corresponding values calculated by assuming two and four times better point source sensitivity are shown in parentheses.}
\label{tab:numbers}
\end{center}
\end{table}

\begin{table}[t]
\begin{center}
\begin{tabular}{|c||c|c|c|c|c| }
\hline

        & $\alpha_L=1$                 & $\alpha_L=1.5$                  & $\alpha_L=2$  \\
\hline \hline
$F(> 10^{-10}$ ph cm$^{-2}$ s$^{-1})$/ $F(>4~10^{-12}$ ph cm$^{-2}$ s$^{-1}$) &   $94\%$           &         $82\%$               &    $52\%$   \\

\hline
\end{tabular}
\caption{Ratio of the cumulative flux of the simulated MSP sample above the flux of $10^{-10}$ ph cm$^{-2}$ s$^{-1}$, to the total cumulative flux calculated above our lowest flux bin, $4\times 10^{-12}$ ph cm$^{-2}$ s$^{-1}$. Numbers are given for the case with $L^{\rm max}=1.5\times 10^{37}$ ph s$^{-1}$. }
\label{tab:ratioflux}
\end{center}
\end{table}

It is important to note that at fluxes above \fermi LAT threshold one is probing the end-point of the luminosity function, where a 50\% change in $L^{\rm max}$ can easily translate into an order of magnitude change in the expected number of detected sources. {Even {\it assuming} zero MSPs detected by \fermi LAT}, we remind the reader that an expected number of events as large as 3 would still be compatible at 95\% with a lack of observations.  {Note that at the moment it is unclear if/how many MSPs from this region of the sky have been detected, yet. This is a tricky problem since the definitions of a source {\it detection} and its {\it identification} as belonging to a particular class of objects may differ. In the case of the 2PC catalog used for this study, additional requirements are clearly demanded to enter the catalog, such as detection of a {\it pulsation} from the source. The numbers obtained in our table are in this respect optimistic, representing a number of spectrally identified `pulsar like' candidate sources \fermi LAT could observe in the ROI.
To illustrate this point, it is worth mentioning that  about $60$ spectrally tagged ``pulsar-like'' candidates were found within $2.5^\circ$ of the Galactic plane and about $30$ at intermediate latitudes, $2.5^\circ \lsi |b|\lsi 10^\circ$ already in 1FGL \cite{Fermi-LAT:2011sla}.  Intriguingly, preliminary results from the 3FGL catalog analysis~\footnote{http://fermi.gsfc.nasa.gov/science/mtgs/symposia/2014/program/14A$\_$Saz$\_$Parkinson.pdf} { had already shown} several tens of unassociated sources within 10$^\circ$ from the GC with spectral properties consistent with being MSPs, { while the recently published 3FGL catalog \cite{TheFermi-LAT:2015hja} finds seven point sources within $1^\circ$ from the Galactic Center, although they still refrain from a detailed quantitative analysis of the region, due to its complexity}. Actual comparisons with experiment would  require computing the number of MSP-like events expected to be seen in a given ROI, based on the different luminosity functions considered above, for a given set of selection cuts, sensitivity threshold, etc. applied to the instrument, which is only possible to be done properly within the collaboration. Nonetheless, we believe that our results are already very interesting: we see from the Table \ref{tab:numbers} that for a low value of $\LM$,  MSPs can attain the level of the observed residuals while remaining still compatible with zero MSP sources from the bulge, independently of the value of $\alpha_L$ used.  For sufficiently large  $\alpha_L$, even values of $\LM$ a bit larger can be tolerated without the lack of observations generating troubles.}

Actually, even within the simplified model above, it may be even easier to accommodate the two apparently conflicting observations since:
\begin{itemize}
\item {\it Source confusion} may play a role in degrading sensitivity: the point source sensitivity is usually estimated for an isolated source against a diffuse background and known point-sources. Since the diffuse background is higher along the Galactic Plane and in particular in the inner Galaxy, it is not surprising that a lower sensitivity is found in that region, see for instance \cite{Gregoire:2013yta,TheFermi-LAT:2013ssa}. However, the inner Galaxy region is expected to be overcrowded, with a concrete possibility that several sources might be spatially overlapping (we remind that at $\sim1\,$GeV energy, where MSP emission peaks, the angular resolution is of the order of a degree). The effective point source sensitivity for densely packed sources with MSPs-like spectra might be thus lower than what we assumed. In practice, the border between detection and non-detection of a point-source may depend crucially on the fewer energetic photons populating the exponentially suppressed cutoff of the MSP spectrum, which can be reconstructed with sufficient angular accuracy. This is critical since small changes in the LAT point source sensitivity lead to a strongly varying number for the predictions.
\item  An additional handle is provided by the GCE flux that needs to be accounted for in the first place. Current uncertainties reported in Table~\ref{tab2} could lower the need of a source for the GCE by  
{\it a factor two} with respect to the fiducial value. This implies that a number of resolved sources roughly a factor of two lower than what reported in Tab.~\ref{tab:numbers} might still be consistent with explaining the totality of the observed GCE, within the uncertainties.
\end{itemize}

{Nonetheless, it is clear that our conclusions crucially depend on the choice for $L^{\rm max}$ and (to lesser extent) for $\alpha_L$. Too large values $\alpha_L\sim 2$, for instance, might lead to overshoot the number of sources at lower luminosities, if this luminosity function was extrapolated down over few orders of magnitudes. But this does not appear to be a problem, { since viable solutions also exist with $\alpha_L\sim 1\-- 1.5$,} or may 
also tolerate $\alpha_L\sim 2$ with a break to a smaller index at low luminosities, for instance. {The usefulness of performing a parametric study is particularly evident for $L^{\rm max}$: we clearly obtain a range
of predictions ranging from results qualitatively similar to Ref.~\cite{Cholis:2014noa}, where the scarcity of resolved sources would indicate a tension with a MSP solution to the GCE, to perfectly viable solutions when lower
values of $L^{\rm max}$ (yet consistent with current observations) are used. Our results should be considered as a proof of principle that working solutions only relying on prompt emission from
MSP do exist. Too strong conclusions  found in past literature {(e.g. that no MSP explanation is viable~\cite{Hooper:2013nhl})} thus depend on some extrapolations {(e.g. that the luminosity function extends to sufficiently high $L^{\rm max}$)} or on imposing some ``theoretical prejudices'' {(such as an unbroken power-law form of  the luminosity function)}, and are in general not robust with respect to lifting these hypotheses.}
An encouraging conclusion of our study is that, while viable values of $\LM$ exist for which MSPs can both be consistent with {\it Fermi} LAT (paucity or) lack of detections of bulge MSPs and account for the GCE with their unresolved component, a two to four times better point source sensitivity should be sufficient in all cases to start resolving a significant number of MSPs in the bulge (see Table \ref{tab:numbers}) and therefore support or disprove the baseline hypothesis.

 {Of course, modifications of the simplified framework can alter somewhat the conclusions, a point we turn now to discuss in some detail. We already mentioned the ``obvious'' effect of reducing the statistics of detected sources by introducing a break in the luminosity function or allowing for a variations in the spectrum.}   {More in general, it is clear that in order to to relax potential tensions between GCE explanations and resolved MSP statistics at the GC one can invoke the existence of} {\it significantly different populations} of MSPs~\cite{Story:2007xy,2011ApJ...743..181H}. Although the properties of MSPs deduced from the local neighborhood and globular clusters are roughly compatible, one can certainly wonder if some features of the MSP populations are set by the environmental conditions and/or MSP formation history, causing a scatter among best fit parameters in different regions. A more extended discussion of this possibility can be found in~\cite{Mirabal:2013rba}. In \cite{Cholis:2014noa} it was observed that the average spectrum of MSPs in the MW disk differs from the spectra of the MSPs in globular clusters, provided that there are not other unresolved source populations contributing to the  globular clusters emission, of course. In addition, although the MW and the globular cluster population luminosity functions appear similar at medium and low luminosities, the shape close to the high luminosity end cannot be probed in globular clusters (due to their larger distance), so no comparison of $L_{\rm max}$ can be made between these two regions.   From the theoretical point of view, there might be reasons for a difference between bulge and disk populations: it is well known that the stellar population in globular clusters is older than the average stellar population in the GC, or that the environmental  gas density is significantly higher in the GC. For instance, it is reasonable that the expected differences in the birth rate (or mass ratios) of binary systems at the GC vs. the local neighborhood may have an impact on MSP populations, too.

As a further example, we mention the possibility that the MSP emission currently accounted for  may not be the whole story, and that secondary emission might play a major role.  We devote the next section to more details on this {new} idea.

\section{Inverse Compton emission from MSPs}\label{ICmsps}

It has been argued that the efficiency of gamma ray emission of MSPs, \ie ratio of gamma ray luminosity to total spin down rate,  is near $10\%$, as opposed to efficiencies on the order of a few percent for non-recycled, ordinary pulsars~\cite{Story:2007xy,Johnson:2012rk,Venter:2014zea}. By analogy with ordinary pulsars, for which the wind nebula emission is measured, this suggests that there might be roughly one order of magnitude more energy emitted in a relativistic wind of  $e^\pm$ with respect to the gamma ray signal.

The MSP {\it beaming factors} are often assumed to be $\sim 1$, implying that nearly all pulsars contribute to gamma ray signals observable at Earth. However, the study~\cite{Pierbattista:2012nn} found a large spread of the beaming factors among the different emission models and, in more general terms, between radio-loud and radio-quiet pulsars. Accordingly, it is not unlikely that a higher fraction of MSP than those visible in gamma-rays may actually contribute to releasing relativistic pairs, with overall energy carried by these particles possibly one order of magnitude larger than the one visible in prompt gammas.

Electrons with energy in the 10 GeV energy range lose most of their energy via Inverse Compton (IC) on CMB and interstellar radiation, yielding up-scattered gamma-ray photons, as well as synchrotron emission, which falls in the radio-band. It is tempting to compute the expected IC gamma ray signal coming from the electron population of the MSPs and compare it with the prompt emission from MSPs and the observed excess at the GC.

There is little knowledge on the actual spectra of relativistic leptons released by MSPs in the interstellar medium. Hence we decide to conduct a phenomenological study, injecting an electron/positron population with power law with exponential cut-off spectra, \ie
\begin{equation}
\frac{d\Phi_{e}}{dE}= K \left( \frac{E}{E_{0,e}}\right)^{-\Gamma_{e}} \exp\left(-\frac{E}{E_{\rm cut, e}}\right)\,. \label{eq:elspectra}
\end{equation}

 Concerning the index and the cut-off energy,  we test several different choices, reported in Table~\ref{tab:IC}. 

\begin{table}[t]
\begin{center}
\begin{tabular}{|c|c| }
\hline

\textbf{Parameter}  &{  \textbf{Values} } \\
$\Gamma_{e}$&    1.5, 2.2 \\
E$_{\rm cut, ~e}$ [GeV]&   3, 30, 300 \\
\hline
\end{tabular}
\caption{Values of $\Gamma_e$ and $E_{\rm cut,~e}$ describing the electron-positron spectra from MSPs, see Eq.~(\ref{eq:elspectra}).}
\label{tab:IC}
\end{center}
\end{table}

As for the relative normalization of the photon and total electron fluxes, we fix $K$ in Eq.~(\ref{eq:elspectra}) by requiring an {\it equal} total energy stored in the two species, \ie 
 \begin{equation}
\int _{\geq {\rm 1 GeV}}  E \frac{d\Phi_{e}}{dE} dE = \int _{\geq {\rm 1 GeV}} E \frac{d\Phi_{\gamma}}{dE} dE\,.\label{normepem}
 \end{equation}
For the spatial distribution of MSPs we follow the same prescription we used in the previous sections, Eq.~(\ref{eq:rho}), with parameters from Table \ref{tab1}. We also tested
the situation in which all the injected pairs are injected with a Gaussian distribution relatively near the GC within 100 pc, and in a broader region of 1 and 3 kpc distance. Here we assume ``standard'' values for cosmic ray propagation parameters: diffusion coefficient at 4 GV of $10^{29}$ cm$^2$s$^{-1}$, diffusion index of 0.33, reacceleration speed of 30 kms$^{-1}$, halo height of 10 kpc and no convection (for the definition of parameters and values consistent with cosmic ray and gamma ray measurements see \eg \cite{FermiLAT:2012aa}).}
For the magnetic field (responsible for the energy loss via synchrotron, not visible in the gamma band) we make a standard choice of $5 ~\mu$G, but we checked that even adopting $B=20\mu$G would only alter lower the results by a factor $\sim 2$. Also, we checked that with the possible exception of the inner degree region bremsstrahlung emission is subdominant, essentially for the same reasons reported in~\cite{Petrovic:2014uda}. 

The computation of the spectra and morphology of the signal was eventually made via the {\sc GALPROP} code, whose DM routine we modified to inject prompt photons, as well as electrons and positrons, with the MSP like spectra of Eq.~(\ref{eq:spectra}) and Eq.~(\ref{eq:elspectra}), respectively. 

In the top panel of Fig.~\ref{fig:IC} we show our results for the morphology of the secondary emission at the energy $E=1\,$GeV, with arbitrary normalization, 
while in the bottom panel we show energy spectra averaged in the region $|b,l|\leq 20^\circ$ and $|b|\geq 2^\circ$, with the relative normalization given by Eq.~(\ref{normepem}). In both cases, we report the prompt emission normalized to fluxes found in \cite{Calore:2014xka}, for comparison.

Qualitatively, the top panel shows that the profile of the GCE can be mimicked quite closely at intermediate and high
latitudes by a lepton injection concentrated in the inner Galaxy. Actually, we checked that choosing an injection region smaller than 100 pc (or a bit larger) would
not change the morphology, which is dominated by diffusion-losses.  It is remarkable that the morphology of the signal between $\sim 3^\circ$ and $\sim 12^\circ$
is {\it naturally expected} for a signal whose origin is an IC process from electrons injected near the GC, a property noted in~\cite{Petrovic:2014uda}.
The red/dot-dashed curve and orange/dashed curves show that by enlarging the injection region the morphology flattens out.  Even if we take an injection going as $r^{-2.4}$, 
as suggested by a prompt component fit, the secondary emission results {flattens out}, which suggests that the ratio of secondary/primary emission should be a growing function of
the distance.  
In all cases,  the IC emission does not appear to be cuspy enough to
match the morphology of the GCE in the inner few degrees. We warn however the reader that this conclusion is not necessarily
robust with respect to variations in the model parameters: for instance, different propagation parameters
in the inner region of the Galaxy might at least partially account for the difference, see also the discussion and results in~\cite{Petrovic:2014uda}. 

The bottom panel of Fig.~\ref{fig:IC} illustrates another couple of interesting points:
\begin{itemize}
\item Following the ``conservative'' normalization given by Eq.~(\ref{normepem}),  it is very reasonable to expect a secondary contribution at the 10\%-20\% level of the prompt emission, unless the cutoff is so low that a very small fraction of photons is up-scattered in the relevant energy range. Keeping in mind that there is in principle one order of magnitude larger energy reservoir in MSPs, based on energetics arguments alone one cannot exclude that 
secondary emission contributes to a significant fraction of the total GCE. 
\item  If the maximal energy of the electrons is comparable to the maximal energy of prompt gamma rays, not surprisingly secondary emission falls short of explaining
the high-energy end of the GCE spectrum. In order to match the spectral endpoint, one needs electrons to be accelerated to several tens of GeV. Furthermore,  if the electrons are accelerated to significantly higher energies, say up to hundreds of GeV, one expects the spectrum of the GCE to extend to higher energies than the few GeV typical of prompt emission of MSPs. This would thus be visible even if the lower energy emission is dominated by prompt photons. Actually, according to the analysis of \cite{Calore:2014xka} there is some hint for such an extended dynamical range. 
\end{itemize}
Note that clarifying this feature would be extremely important for diagnostics: in DM models, the maximal energy of secondary photons is basically always below the one of prompt photons. For pulsars, it needs not to be the case: For ordinary pulsars surrounded by pulsar wind nebulae, one can in fact infer the presence of hundreds GeV to TeV electrons at the termination shock from radio and X rays, attributed to extra acceleration well beyond the light cylinder~\cite{Blasi:2010de}.  An indirect hint that such high energies are reached in pulsar complexes is linked to the ``positron excess'' extending to hundreds of GeV measured by PAMELA, \fermi LAT and AMS, for which these objects constitute the most likely
explanation (see for instance the review~\cite{Serpico:2011wg}.)~\footnote{For a first attempt at estimating the MSP contribution to local electron-positron fluxes, see the recent~\cite{Venter:2014ata}.}. 
Unfortunately, it is unclear if the environments associated to MSPs provide an extra acceleration mechanism for electrons. 
Since MSPs are not surrounded by nebulae, one could naively assume that the pair spectra emerging from the MSPs are good representations of the intrinsic source spectra. But, just to name a possible counter-argument, it is known that the majority of MSPs are in binary systems: a sub-sample of these can further accelerate the pairs at strong inter-binary shocks. This is why we left  $E_{\rm max}$ as a free parameter in the preliminary parametric study above. 

In summary, the energy stored in secondary emission from MSP is likely important at least at the 10\% level of the GCE, and could conceivably account for a larger fraction of the signal~\footnote{A recent paper which analyzed the \fermi LAT data found intriguing correlations of the GCE with micron infrared emission morphology~\cite{Abazajian:2014hsa}. This  would suggest that secondary emission from electrons (as opposed to prompt emission) plays a crucial role, consistently with our findings here.}.  
Our calculation also suggests that it has roughly the right angular profile at intermediate and large scales, while in the inner degrees it is harder to predict firmly. Reproducing the spectrum requires additionally some extra assumption on the maximal acceleration of electrons in these objects, which is poorly known. If secondary emission is to provide a dominant fraction of the GCE signal, it is not so obvious to what extent the quasi-universality of the GCE spectrum as a function of Galactic coordinates can be reproduced
without fine-tuning of the parameters. On one hand, it is more natural to expect variations in the spectra at the few tens of percent, especially at low energies, see~\cite{Petrovic:2014uda}. On the other hand, if the secondary contribution is sub-leading, an intriguing signature of the MSP model may be the emergence of an extended spectral component at tens of GeV, provided that electron acceleration up to hundreds of GeV takes place in these objects.
 
Inverse Compton signals would naturally have counterparts at radio wavelengths. Studies of radio emission counterpart of  the GCE within dark matter frameworks have been done in \cite{Bringmann:2014lpa,Cholis:2014fja}. They showed high potential to constrain the electron population, but also emphasized high uncertainty in the prediction of the radio signal related to our ignorance on the DM profile or electron propagation parameters in the Galactic environment at the sub parsec scales. Nonetheless, the future SKA telescope array\footnote{https://www.skatelescope.org/} should have  sub-mJy sensitivity and arcsec resolution below 1 GHz, and can be expected to discover both diffuse counterparts of the IC emission and more MSPs as point sources. 

{ Analogously, if the acceleration of primary electrons in MSPs extends to very high-energy, signals may be expected in Cherenkov Telescopes, as it has been argued for instance in~\cite{Bednarek:2013oha}. Current thresholds are typically too high to derive meaningful constraints, but perspectives for the next generation of instruments, notably CTA~\footnote{https://www.cta-observatory.org/}, have recently been studied~\cite{Yuan:2014yda} and appear promising, at least in some part of the parameter space.}
 
\begin{figure}[t]
\begin{center}
\includegraphics[width=0.58\textwidth]{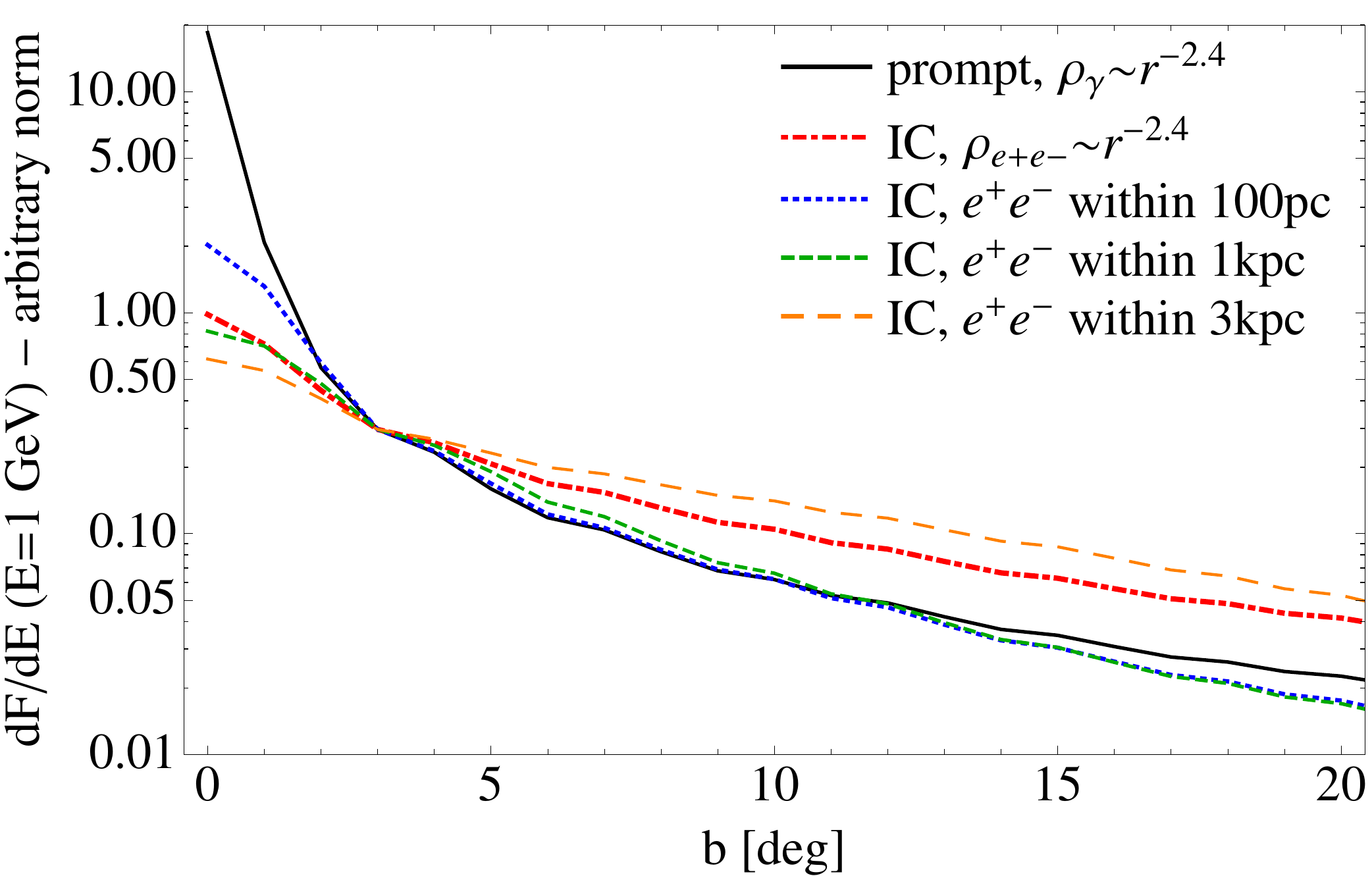}
\includegraphics[width=0.58\textwidth]{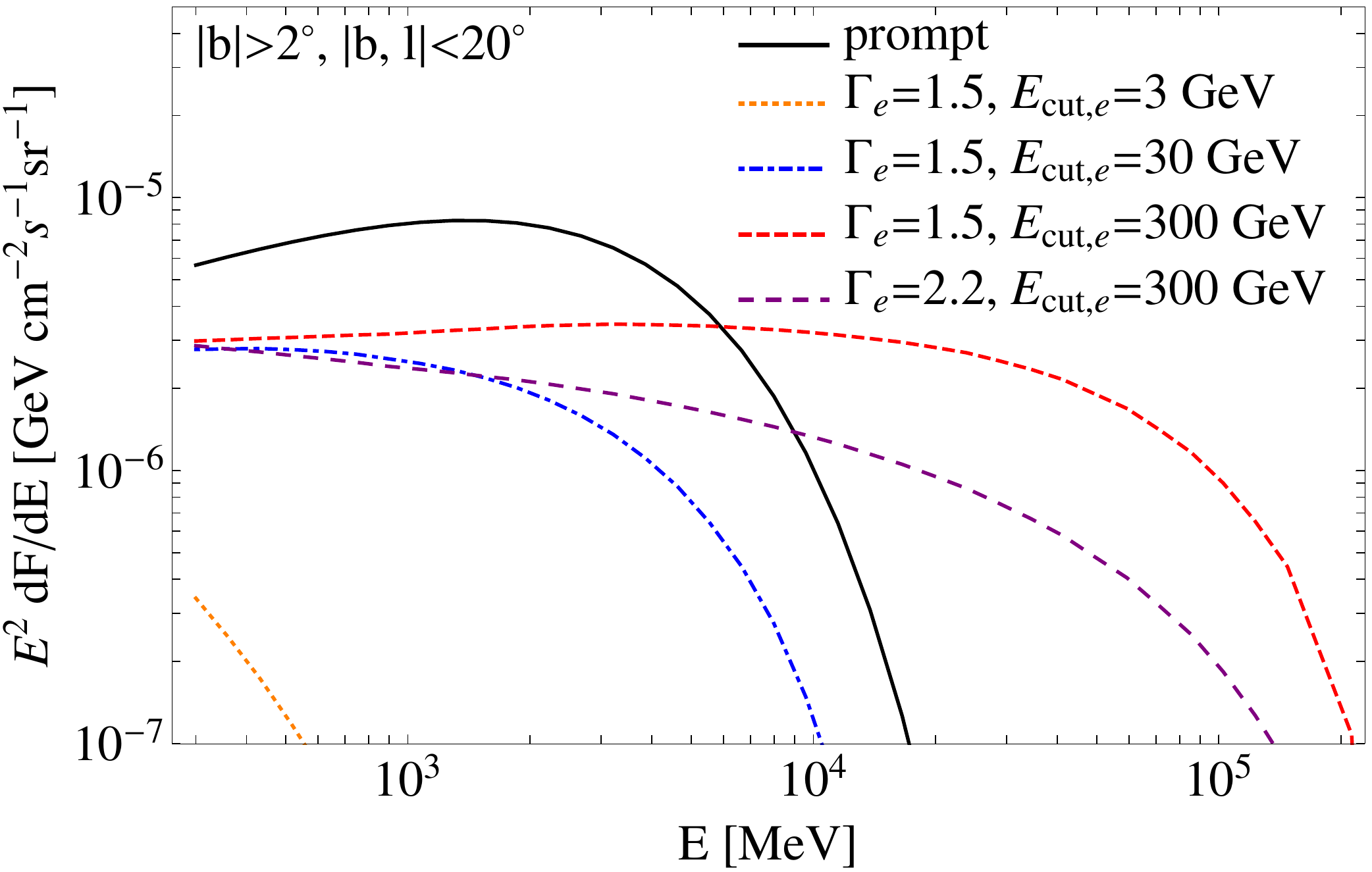}
\caption{{\it Top panel:}  Latitude distribution of the prompt (Black, Solid) and IC (Red, DotDashed) fluxes {at $E=1\,$GeV} produced by MSPs distributed with a distance from the Galactic center falling as $r^{-2.4}$ (normalization is chosen arbitrarily, to match at 4 degrees). We also show the latitude dependence of the IC emission coming from electrons and positrons injected with a Gaussian distribution, with a Gaussian width of 100 pc (Blue, Dotted), 1 kpc (Green, Dashed) and 3 kpc (Orange, Long-dashed) from the GC. {\it Lower panel:} Comparison of the prompt MSP fluxes (Black) with IC fluxes calculated for several choices of the input spectral parameters, listed in Table \ref{tab:IC}, averaged over $|b,l|\leq 20^\circ$ and $|b|\geq 2^\circ$. The normalization of prompt emission is chosen to match fluxes found in \cite{Calore:2014xka}, while relative normalization of the IC emission is set by Eq.~(\ref{normepem}).  \label{fig:IC}}
\end{center}
\end{figure}

\section{Conclusions}
\label{sec:conclusions}

We have revisited the question of the contribution of an unresolved population of millisecond pulsars (MSPs) to the gamma-ray excess from the Galactic Center (GCE). In the recent past, the issue has been raised of the extent to which the signal can be explained by MSPs, while at the same time respecting the bound imposed by the non-detection of resolved sources by \fermi LAT~\cite{Hooper:2013nhl,Yuan:2014rca,Cholis:2014noa,Cholis:2014lta}.
 
To be complementary to previous investigations}, in this article we have performed an analysis which makes a minimum of theoretical assumptions, {adopting a parametric approach and} relying as much as possible on empirical knowledge of the MSP spectra and luminosity {to justify and bracket the relevant ranges of parameters. { For instance, we used the GALPLOT tool, using as a input directly the luminosity function, rather than theoretical computations based on models linking physical parameters (such as the pulsar magnetic field $B$, the hypothesis of pure dipole braking, etc.) to the gamma-ray luminosity, as e.g. in~\cite{Hooper:2013nhl}.}

{Our predictions qualitatively recover the conclusions of } recent works such as~\cite{Cholis:2014lta} when a hard index and a high cutoff value for the luminosity function are assumed. However, we showed that the answer is in fact strongly dependent on the modeling of the end-point of the MSP luminosity function and on the robustness of the \fermi LAT resolution threshold in presence of high-levels of diffuse backgrounds {\it as well as} high density of unresolved sources. We found that within current uncertainties, it is {certainly} possible for unresolved MSP to account for most or all of the GCE without violating any constraints. 
The good news is that, unless the GC population of MSP is significantly different from the others we know, future gamma ray observatories should be able to resolve {a significant number of} MSPs.  Actually, the performances expected by the forthcoming \fermi LAT data analysis (Pass 8) might be sufficient to that purpose: The expected improvement of the point source sensitivity of Pass 8 with respect to currently used Pass 7 is a factor of 1.4 at 1 GeV\footnote{http://fermi.gsfc.nasa.gov/science/mtgs/symposia/2014/program/10B\_Bruel.pdf}. In addition, Pass 8 data will come with four event classes (0 to 3), with an increasingly better PSF, but subsequently smaller statistics due to more severe event cuts needed to reach higher quality levels. As statistics is another critical factor in detection of faint sources, it is still hard to quantify the effective improvement, but it is nonetheless obvious that we should expect a surge in the number of detected sources in the near future, as already hinted by preliminary analyses which uses four years of Pass 7 Reprocessed data\footnote{http://fermi.gsfc.nasa.gov/science/mtgs/symposia/2014/program/02\_Cavazzuti.pdf}. In case of inconclusive answers, however, one may need to
wait for future instruments. In particular, the newly proposed high resolution gamma-ray telescope PANGU \cite{Wu:2014tya}---a joint mission of ESA (European Space Agency) and CSA (Chinese Space Agency)---should have an unprecedented point spread function thanks to an innovative micro-strip detector of silicon layers, as well as a spectrometer that specifically targets the sub-GeV region (focusing on the $\gsi 10$ MeV to $\lsi$ 1 GeV). 
It is expected that PANGU will have $\gsi$ 5 times better PSF then \fermi  LAT, thus separating point sources from the diffuse background with high precision.
 
An additional opportunity may be provided by the confirmed new satellite mission ``Gamma Astronomical Multifunctional Modular Apparatus'' - GAMMA-400 telescope  \cite{Galper:2013sfa}, scheduled  to fly in  2018/19.  In the baseline configuration, GAMMA-400 will be optimized for the energy of  100 GeV with angular resolution ~0.01$^\circ$ and energy resolution $\sim 1\%$, both an order of magnitude better than the Fermi LAT. At $E\sim$1 GeV its performance will be comparable to the \fermi LAT and thus not necessarily resolutive. However, among the currently proposed configurations, there is also a possibility for an upgraded Si tracker, which would improve angular resolution by a factor of few at 1 GeV and therefore directly address the possible point sources origin of the GCE on a relatively short timescale \footnote{See \eg {\href{http://www.ba.infn.it/~now/now2012/web-content/TALKS/Monday10/parallel2/NOW2012_Bonvicini.pptx}{talk}} by V. Bonvizini at NOW2012.}. { Needless to say, these improvements will be less decisive should the major obstacle be due to
the complexity in modeling the truly diffuse emission.}

{We summarize below the main results of our paper:
\begin{itemize}
\item We showed that, starting from a parametric study of the prompt emission from the unresolved population of MSPs, with range of parameters wide enough and inspired by observational input and basic theory ---MSPs may account even for the totality of the signal, without violating constraints from lack of resolved sources by \fermi LAT. 
\item We identified what are the most relevant uncertainties to this effect (endpoint of luminosity function, sensitivity of the detector, importance of accounting for errors in the GCE normalization) and estimated the number of detected MSPs in future gamma-ray missions, given these uncertainties.  
\item We further argued that secondary emission (via IC of $e^\pm$) can account for at least a fraction of the GCE, a point that deserves certainly further study in view of the additional handles it gives on the diagnostics. Also, it naturally reproduces the morphology of the GCE at intermediate and large scales, another intriguing point. 
\end{itemize}
All in all, we believe that it is too early to discard conventional astrophysical sources as explanations of the GCE at GeV energy. Even if MSP are not responsible for the bulk of the GCE, it is expected that they contribute at some non-negligible fraction: a better understanding of their flux will allow to subtract their contribution and  draw more definite conclusions on the existence and properties of any residual GCE. Luckily, for further progress in the field one does not have to wait too long, since a first re-assessment of the nature of the GCE may arrive as soon as the next \fermi LAT data analysis (Pass 8) will become available.}

\section{Acknowledgments}

We are grateful to Andy Strong and Andrea Giuliani for numerous discussions and Andrea Albert, Francesca Calore, Ilias Cholis, Alice Harding, Dan Hooper, Tim Linden, Nestor Mirabal, Pablo Saz Parkinson and Christoph Weniger for comments on the manuscript.


\begin{thebibliography}{10}

\bibitem{Abazajian:2014fta}
K.~N. Abazajian, N.~Canac, S.~Horiuchi, and M.~Kaplinghat, ``{Astrophysical and
  Dark Matter Interpretations of Extended Gamma Ray Emission from the Galactic
  Center},''
\href{http://arxiv.org/abs/1402.4090}{{\ttfamily arXiv:1402.4090
  [astro-ph.HE]}}.

\bibitem{Daylan:2014rsa}
T.~Daylan {\em et~al.}, ``{The Characterization of the Gamma-Ray Signal from
  the Central Milky Way: A Compelling Case for Annihilating Dark Matter},''
\href{http://arxiv.org/abs/1402.6703}{{\ttfamily arXiv:1402.6703
  [astro-ph.HE]}}.

\bibitem{Hooper:2011ti}
D.~Hooper and T.~Linden, ``{On The Origin Of The Gamma Rays From The Galactic
  Center},'' \href{http://dx.doi.org/10.1103/PhysRevD.84.123005}{{\em
  Phys.Rev.} {\bfseries D84} (2011) 123005},
\href{http://arxiv.org/abs/1110.0006}{{\ttfamily arXiv:1110.0006
  [astro-ph.HE]}}.

\bibitem{Hooper:2010mq}
D.~Hooper and L.~Goodenough, ``{Dark Matter Annihilation in The Galactic Center
  As Seen by the Fermi Gamma Ray Space Telescope},''
  \href{http://dx.doi.org/10.1016/j.physletb.2011.02.029}{{\em Phys.Lett.}
  {\bfseries B697} (2011) 412--428},
\href{http://arxiv.org/abs/1010.2752}{{\ttfamily arXiv:1010.2752 [hep-ph]}}.

\bibitem{Hooper:2013rwa}
D.~Hooper and T.~R. Slatyer, ``{Two Emission Mechanisms in the Fermi Bubbles: A
  Possible Signal of Annihilating Dark Matter},''
  \href{http://dx.doi.org/10.1016/j.dark.2013.06.003}{{\em Phys.Dark Univ.}
  {\bfseries 2} (2013) 118--138},
\href{http://arxiv.org/abs/1302.6589}{{\ttfamily arXiv:1302.6589
  [astro-ph.HE]}}.

\bibitem{Boyarsky:2010dr}
A.~Boyarsky, D.~Malyshev, and O.~Ruchayskiy, ``{A comment on the emission from
  the Galactic Center as seen by the Fermi telescope},''
  \href{http://dx.doi.org/10.1016/j.physletb.2011.10.014}{{\em Phys.Lett.}
  {\bfseries B705} (2011) 165--169},
\href{http://arxiv.org/abs/1012.5839}{{\ttfamily arXiv:1012.5839 [hep-ph]}}.

\bibitem{Abazajian:2012pn}
K.~N. Abazajian and M.~Kaplinghat, ``{Detection of a Gamma-Ray Source in the
  Galactic Center Consistent with Extended Emission from Dark Matter
  Annihilation and Concentrated Astrophysical Emission},''
  \href{http://dx.doi.org/10.1103/PhysRevD.86.083511}{{\em Phys.Rev.}
  {\bfseries D86} (2012) 083511},
\href{http://arxiv.org/abs/1207.6047}{{\ttfamily arXiv:1207.6047
  [astro-ph.HE]}}.

\bibitem{Gordon:2013vta}
C.~Gordon and O.~Macias, ``{Dark Matter and Pulsar Model Constraints from
  Galactic Center Fermi-LAT Gamma Ray Observations},''
  \href{http://dx.doi.org/10.1103/PhysRevD.88.083521}{{\em Phys.Rev.}
  {\bfseries D88} (2013) 083521},
\href{http://arxiv.org/abs/1306.5725}{{\ttfamily arXiv:1306.5725
  [astro-ph.HE]}}.

\bibitem{Calore:2014xka}
F.~Calore, I.~Cholis, and C.~Weniger, ``{Background model systematics for the
  Fermi GeV excess},''
\href{http://arxiv.org/abs/1409.0042}{{\ttfamily arXiv:1409.0042
  [astro-ph.CO]}}.

\bibitem{Vitale:2009hr}
{\bfseries Fermi/LAT Collaboration}, V.~Vitale and A.~Morselli, ``{Indirect
  Search for Dark Matter from the center of the Milky Way with the Fermi-Large
  Area Telescope},''
\href{http://arxiv.org/abs/0912.3828}{{\ttfamily arXiv:0912.3828
  [astro-ph.HE]}}.

\bibitem{Huang:2013apa}
W.-C. Huang, A.~Urbano, and W.~Xue, ``{Fermi Bubbles under Dark Matter Scrutiny
  Part II: Particle Physics Analysis},''
  \href{http://dx.doi.org/10.1088/1475-7516/2014/04/020}{{\em JCAP} {\bfseries
  1404} (2014) 020},
\href{http://arxiv.org/abs/1310.7609}{{\ttfamily arXiv:1310.7609 [hep-ph]}}.

\bibitem{Gaggero:2014xla}
D.~Gaggero, A.~Urbano, M.~Valli, and P.~Ullio, ``{The gamma-ray sky points to
  radial gradients in cosmic-ray transport},''
\href{http://arxiv.org/abs/1411.7623}{{\ttfamily arXiv:1411.7623
  [astro-ph.HE]}}.

\bibitem{Mirabal:2014ila}
N.~Mirabal, ``{Annihilating dark matter or noise?: A statistical examination of
  the Fermi GeV excess around the Galactic Centre},''
\href{http://arxiv.org/abs/1411.7410}{{\ttfamily arXiv:1411.7410
  [astro-ph.HE]}}.

\bibitem{Yuan:2014rca}
Q.~Yuan and B.~Zhang, ``{Millisecond pulsar interpretation of the Galactic
  center gamma-ray excess},''
\href{http://arxiv.org/abs/1404.2318}{{\ttfamily arXiv:1404.2318
  [astro-ph.HE]}}.

\bibitem{Mirabal:2013rba}
N.~Mirabal, ``{Dark matter vs. Pulsars: Catching the impostor},'' {\em MNRAS}
  {\bfseries 436} (2013) 2461,
\href{http://arxiv.org/abs/1309.3428}{{\ttfamily arXiv:1309.3428
  [astro-ph.HE]}}.

\bibitem{Carlson:2014cwa}
E.~Carlson and S.~Profumo, ``{Cosmic Ray Protons in the Inner Galaxy and the
  Galactic Center Gamma-Ray Excess},''
\href{http://arxiv.org/abs/1405.7685}{{\ttfamily arXiv:1405.7685
  [astro-ph.HE]}}.

\bibitem{Petrovic:2014uda}
J.~Petrovic, P.~D. Serpico, and G.~Zaharijas, ``{Galactic Center gamma-ray
  "excess" from an active past of the Galactic Centre?},''
\href{http://arxiv.org/abs/1405.7928}{{\ttfamily arXiv:1405.7928
  [astro-ph.HE]}}.

\bibitem{Ponti:2012pn}
G.~Ponti, M.~R. Morris, R.~Terrier, and A.~Goldwurm, ``{Traces of past activity
  in the Galactic Centre},''
  \href{http://dx.doi.org/10.1007/978-3-642-35410-6_26}{{\em Astrophys.Space
  Sci.Proc.} {\bfseries 34} (2013) 331--369},
\href{http://arxiv.org/abs/1210.3034}{{\ttfamily arXiv:1210.3034
  [astro-ph.GA]}}.

\bibitem{Hooper:2013nhl}
D.~Hooper, I.~Cholis, T.~Linden, J.~Siegal-Gaskins, and T.~Slatyer,
  ``{Millisecond pulsars Cannot Account for the Inner Galaxy's GeV Excess},''
  \href{http://dx.doi.org/10.1103/PhysRevD.88.083009}{{\em Phys.Rev.}
  {\bfseries D88} (2013) 083009},
\href{http://arxiv.org/abs/1305.0830}{{\ttfamily arXiv:1305.0830
  [astro-ph.HE]}}.

\bibitem{Cholis:2014lta}
I.~Cholis, D.~Hooper, and T.~Linden, ``{Challenges in Explaining the Galactic
  Center Gamma-Ray Excess with Millisecond Pulsars},''
\href{http://arxiv.org/abs/1407.5625}{{\ttfamily arXiv:1407.5625
  [astro-ph.HE]}}.

\bibitem{Cholis:2014noa}
I.~Cholis, D.~Hooper, and T.~Linden, ``{A New Determination of the Spectra and
  Luminosity Function of Gamma-Ray Millisecond Pulsars},''
\href{http://arxiv.org/abs/1407.5583}{{\ttfamily arXiv:1407.5583
  [astro-ph.HE]}}.

\bibitem{TheFermi-LAT:2013ssa}
{\bfseries The Fermi-LAT collaboration}, A.~Abdo {\em et~al.}, ``{The Second
  Fermi Large Area Telescope Catalog of Gamma-ray Pulsars},''
  \href{http://dx.doi.org/10.1088/0067-0049/208/2/17}{{\em Astrophys.J.Suppl.}
  {\bfseries 208} (2013) 17},
\href{http://arxiv.org/abs/1305.4385}{{\ttfamily arXiv:1305.4385
  [astro-ph.HE]}}.

\bibitem{Johnson:2012rk}
T.~Johnson, ``{Constraints on the Emission Geometries of Gamma-ray Millisecond
  Pulsars Observed with the Fermi Large Area Telescope},''
\href{http://arxiv.org/abs/1209.4000}{{\ttfamily arXiv:1209.4000
  [astro-ph.HE]}}.

\bibitem{Caraveo:2013lra}
P.~A. Caraveo, ``{Gamma-ray Pulsar Revolution},''
\href{http://arxiv.org/abs/1312.2913}{{\ttfamily arXiv:1312.2913
  [astro-ph.HE]}}.

\bibitem{Gregoire:2013yta}
T.~Gregoire and J.~Knodlseder, ``{Constraining the Galactic millisecond pulsar
  population using Fermi Large Area Telescope},''
\href{http://arxiv.org/abs/1305.1584}{{\ttfamily arXiv:1305.1584
  [astro-ph.GA]}}.

\bibitem{2010ApJS..187..460A}
A.~A. {Abdo} {\em et~al.}, ``{The First Fermi Large Area Telescope Catalog of
  Gamma-ray Pulsars},''
  \href{http://dx.doi.org/10.1088/0067-0049/187/2/460}{{\em Astrophys.J.Suppl.}
  {\bfseries 187} (Apr., 2010) 460--494},
  \href{http://arxiv.org/abs/0910.1608}{{\ttfamily arXiv:0910.1608
  [astro-ph.HE]}}.

\bibitem{Venter:2014zea}
C.~Venter, T.~Johnson, A.~Harding, and J.~Grove, ``{Modelling the light curves
  of Fermi LAT millisecond pulsars},''
\href{http://arxiv.org/abs/1411.0559}{{\ttfamily arXiv:1411.0559
  [astro-ph.HE]}}.

\bibitem{Strong:2006hf}
A.~W. Strong, ``{Source population synthesis and the Galactic diffuse gamma-ray
  emission},'' \href{http://dx.doi.org/10.1007/s10509-007-9480-1}{{\em
  Astrophys.Space Sci.} {\bfseries 309} (2007) 35--41},
\href{http://arxiv.org/abs/astro-ph/0609359}{{\ttfamily arXiv:astro-ph/0609359
  [astro-ph]}}.

\bibitem{Verbiest:2014xea}
J.~Verbiest and D.~Lorimer, ``{Why the distance of PSR J0218+4232 does not
  challenge pulsar emission theories},''
\href{http://arxiv.org/abs/1408.0281}{{\ttfamily arXiv:1408.0281
  [astro-ph.HE]}}.

\bibitem{Calore:2014oga}
F.~Calore, M.~Di~Mauro, and F.~Donato, ``{Diffuse gamma-ray emission from
  galactic Millisecond Pulsars},''
\href{http://arxiv.org/abs/1406.2706}{{\ttfamily arXiv:1406.2706
  [astro-ph.HE]}}.

\bibitem{Schnitzeler:2012jq}
D.~Schnitzeler, ``{Modelling the Galactic distribution of free electrons},''
  \href{http://dx.doi.org/10.1111/j.1365-2966.2012.21869.x}{{\em
  Mon.Not.Roy.Astron.Soc.} {\bfseries 427} (2012) 664 0678},
\href{http://arxiv.org/abs/1208.3045}{{\ttfamily arXiv:1208.3045
  [astro-ph.GA]}}.

\bibitem{Hou:2014haa}
{\bfseries Fermi-LAT Collaboration}, X.~Hou {\em et~al.}, ``{Six Faint
  Gamma-ray Pulsars Seen with the Fermi Large Area Telescope -- Towards a
  Sample Blending into the Background},''
  \href{http://dx.doi.org/10.1051/0004-6361/201424294}{{\em Astron.Astrophys.}
  {\bfseries 570} (2014) A44},
\href{http://arxiv.org/abs/1407.6271}{{\ttfamily arXiv:1407.6271
  [astro-ph.HE]}}.

\bibitem{2014AAS...22325603G}
J.~E. {Grove} and {Fermi LAT Collaboration}, ``{Pass 8: Transforming the
  Scientific Performance of the Fermi Large Area Telescope},'' {\em American
  Astronomical Society Meeting Abstracts \#223} {\bfseries 223} (Jan., 2014)
  256.03.

\bibitem{2011CoPhC.182.1156V}
A.~E. {Vladimirov} {\em et~al.}, ``{GALPROP WebRun: An internet-based service
  for calculating galactic cosmic ray propagation and associated photon
  emissions},'' \href{http://dx.doi.org/10.1016/j.cpc.2011.01.017}{{\em
  Computer Physics Communications} {\bfseries 182} (May, 2011) 1156--1161},
  \href{http://arxiv.org/abs/1008.3642}{{\ttfamily arXiv:1008.3642
  [astro-ph.HE]}}.

\bibitem{Wu:2014tya}
X.~Wu {\em et~al.}, ``{PANGU: A High Resolution Gamma-ray Space Telescope},''
\href{http://arxiv.org/abs/1407.0710}{{\ttfamily arXiv:1407.0710
  [astro-ph.IM]}}.

\bibitem{Galper:2013sfa}
A.~Galper {\em et~al.}, ``{The Space-Based Gamma-Ray Telescope GAMMA-400 and
  Its Scientific Goals},''
\href{http://arxiv.org/abs/1306.6175}{{\ttfamily arXiv:1306.6175
  [astro-ph.IM]}}.

\bibitem{Fermi-LAT:2011sla}
{\bfseries Fermi-LAT Collaboration}, ``{A Statistical Approach to Recognizing
  Source Classes for Unassociated Sources in the First Fermi-LAT Catalog},''
  \href{http://dx.doi.org/10.1088/0004-637X/753/1/83}{{\em Astrophys.J.}
  {\bfseries 753} (2012) 83},
\href{http://arxiv.org/abs/1108.1202}{{\ttfamily arXiv:1108.1202
  [astro-ph.HE]}}.

\bibitem{TheFermi-LAT:2015hja}
{\bfseries The Fermi-LAT Collaboration}, ``{Fermi Large Area Telescope Third
  Source Catalog},''
\href{http://arxiv.org/abs/1501.02003}{{\ttfamily arXiv:1501.02003
  [astro-ph.HE]}}.

\bibitem{Story:2007xy}
S.~A. Story, P.~L. Gonthier, and A.~K. Harding, ``{Population synthesis of
  radio and gamma-ray millisecond pulsars from the Galactic disk},''
  \href{http://dx.doi.org/10.1086/521016}{{\em Astrophys.J.} {\bfseries 671}
  (2007) 713--726},
\href{http://arxiv.org/abs/0706.3041}{{\ttfamily arXiv:0706.3041 [astro-ph]}}.

\bibitem{2011ApJ...743..181H}
A.~K. {Harding} and A.~G. {Muslimov}, ``{Pulsar Pair Cascades in Magnetic
  Fields with Offset Polar Caps},''
  \href{http://dx.doi.org/10.1088/0004-637X/743/2/181}{{\em Astrophys.J.}
  {\bfseries 743} (Dec., 2011) 181},
  \href{http://arxiv.org/abs/1111.1668}{{\ttfamily arXiv:1111.1668
  [astro-ph.HE]}}.

\bibitem{Pierbattista:2012nn}
M.~Pierbattista, I.~Grenier, A.~Harding, and P.~Gonthier, ``{Constraining
  gamma-ray pulsar gap models with a simulated pulsar population},''
\href{http://arxiv.org/abs/1206.5634}{{\ttfamily arXiv:1206.5634
  [astro-ph.HE]}}.

\bibitem{FermiLAT:2012aa}
{\bfseries Fermi-LAT Collaboration}, ``{Fermi-LAT Observations of the Diffuse
  Gamma-Ray Emission: Implications for Cosmic Rays and the Interstellar
  Medium},'' \href{http://dx.doi.org/10.1088/0004-637X/750/1/3}{{\em
  Astrophys.J.} {\bfseries 750} (2012) 3},
\href{http://arxiv.org/abs/1202.4039}{{\ttfamily arXiv:1202.4039
  [astro-ph.HE]}}.

\bibitem{Blasi:2010de}
P.~Blasi and E.~Amato, ``{Positrons from pulsar winds},''
\href{http://arxiv.org/abs/1007.4745}{{\ttfamily arXiv:1007.4745
  [astro-ph.HE]}}.

\bibitem{Serpico:2011wg}
P.~D. Serpico, ``{Astrophysical models for the origin of the positron
  'excess'},''
  \href{http://dx.doi.org/10.1016/j.astropartphys.2011.08.007}{{\em
  Astropart.Phys.} {\bfseries 39-40} (2012) 2--11},
\href{http://arxiv.org/abs/1108.4827}{{\ttfamily arXiv:1108.4827
  [astro-ph.HE]}}.

\bibitem{Venter:2014ata}
C.~Venter, A.~Kopp, P.~Gonthier, A.~Harding, and I.~Busching, ``{The
  Contribution of Millisecond Pulsars to the Galactic Cosmic-Ray Lepton
  Spectrum},''
\href{http://arxiv.org/abs/1410.6462}{{\ttfamily arXiv:1410.6462
  [astro-ph.HE]}}.

\bibitem{Abazajian:2014hsa}
K.~N. Abazajian, N.~Canac, S.~Horiuchi, M.~Kaplinghat, and A.~Kwa, ``{Discovery
  of a New Galactic Center Excess Consistent with Upscattered Starlight},''
\href{http://arxiv.org/abs/1410.6168}{{\ttfamily arXiv:1410.6168
  [astro-ph.HE]}}.

\bibitem{Bringmann:2014lpa}
T.~Bringmann, M.~Vollmann, and C.~Weniger, ``{Updated cosmic-ray and radio
  constraints on light dark matter: Implications for the GeV gamma-ray excess
  at the Galactic center},''
\href{http://arxiv.org/abs/1406.6027}{{\ttfamily arXiv:1406.6027
  [astro-ph.HE]}}.

\bibitem{Cholis:2014fja}
I.~Cholis, D.~Hooper, and T.~Linden, ``{A Critical Reevaluation of Radio
  Constraints on Annihilating Dark Matter},''
\href{http://arxiv.org/abs/1408.6224}{{\ttfamily arXiv:1408.6224
  [astro-ph.HE]}}.

\bibitem{Bednarek:2013oha}
W.~Bednarek and T.~Sobczak, ``{Gamma-rays from millisecond pulsar population
  within the central stellar cluster in the Galactic Center},''
  \href{http://dx.doi.org/10.1093/mnrasl/slt084}{{\em Mon.Not.Roy.Astron.Soc.}
  {\bfseries 435} (2013) L14},
\href{http://arxiv.org/abs/1306.4760}{{\ttfamily arXiv:1306.4760
  [astro-ph.HE]}}.

\bibitem{Yuan:2014yda}
Q.~Yuan and K.~Ioka, ``{Testing the millisecond pulsar scenario of the Galactic
  center gamma-ray excess with very high energy gamma-rays},''
\href{http://arxiv.org/abs/1411.4363}{{\ttfamily arXiv:1411.4363
  [astro-ph.HE]}}.

\end{thebibliography}

\providecommand{\href}[2]{#2}\begingroup\raggedright\endgroup

\end{document}